\documentclass{llncs}

\usepackage{natbib}
\usepackage{graphicx}

\usepackage{fancyhdr}
\usepackage{lastpage}
 
\begin{document}

\title{A short review and primer on pupillometry in human computer interaction applications}
\author{Oswald Barral \inst{1}}
\institute{Helsinki Institute for Information Technology HIIT, Department of Computer Science, University of Helsinki\\
\email{oswald.barral@hiit.fi},\\
POBox 68, Helsinki, FI-00014 , Finland
}

\maketitle              

\begin{abstract}
The application of psychophysiological signals in human-computer interaction is a growing field with significant potential for future smart personalised systems. Working in this emerging field requires comprehension of an array of physiological signals and analysis techniques. 

Pupillometry has been studied for over a century, but it has just recently started being used in human-computer interaction setups. Traditionally, pupil size has been used as an indicator of cognitive workload and mental effort. However, pupil size has been linked to other cognitive processes as well, ranging from attention to affective processing. We present a short review on the application of pupillometry in human-computer interaction. 

This paper aims to serve as a primer for the novice, enabling rapid familiarisation with the latest core concepts. We put special emphasis on everyday human-computer interface applications to distinguish from the more common clinical or sports uses of psychophysiology.

This paper is an extract from a comprehensive review of the entire field of ambulatory psychophysiology, including 12 similar chapters, plus application guidelines and systematic review. Thus any citation should be made using the following reference:

{\parshape 1 2cm \dimexpr\linewidth-1cm\relax
B. Cowley, M. Filetti, K. Lukander, J. Torniainen, A. Henelius, L. Ahonen, O. Barral, I. Kosunen, T. Valtonen, M. Huotilainen, N. Ravaja, G. Jacucci. \textit{The Psychophysiology Primer: a guide to methods and a broad review with a focus on human-computer interaction}. Foundations and Trends in Human-Computer Interaction, vol. 9, no. 3-4, pp. 150--307, 2016.
\par}


\end{abstract}
%

\section{Introduction}
Pupil size has been studied largely for its value as an indicator of cognitive workload, mental effort, attention, or affective processing. Research in various fields has explored the application of pupil-based metrics in several human-computer interaction (HCI) scenarios. However, the pupil is usually studied under highly controlled conditions, as it is extremely sensitive to external factors such as changes in ambient light. In fact, the pupil dilation that occurs in response to a change in light exposure is much greater than that experienced as a result of cognitive processing \citep{Beatty2000}. Additionally, pupil behaviour is user- and session-dependent. These factors make the use of pupil size and pupil dilation indices as implicit inputs for real-world HCI settings a non-trivial challenge, yet it is an exciting one, in an area that holds great promise.

\section{Background}
Pupil size is regulated by two muscles (the \textit{sphincter pupillae} and the \textit{dilator pupillae}). By dilating and contracting, the pupil controls the amount of light that enters the eye; however, there are fluctuations in pupil size that are not related to the regulation of light entering the eyes. These fluctuations take place on a much smaller scale, not visible to human observers (around 0.5~mm), and are associated with cognitive processes in the brain. This phenomenon has been subject to study for around 150 years now \citep{schiff1875pupille}. Almost a century after the first studies in this area, the field started to grow significantly (around the 1960s) and it piqued the interest of cognitive psychophysiologists aiming at better understanding the functions of the brain and cognition \citep{hess1964pupil}. From the large body of research that then emerged, those cognitive states achieving the greatest acceptance as able to be inferred through pupil size are mental effort and cognitive overload. In a large set of studies, researchers have shown that there is a correlation between pupil size and the level of demands imposed by a cognitive task. That is, the more demanding the cognitive task, the greater the pupil size \citep{Beatty2000}.

More recently, technological advances in the measurement of pupil size have been accompanied by increased interest among HCI researchers in using this metric as an additional communication channel between humans and machines. 

\section{Methods}
In an HCI connection, pupil size is commonly measured through video-oculography (VOG), a technique using cameras (often infrared cameras) that record the eye and, by applying image processing, allow one not only to track the pupil but also to track the point of regard (the point at which the eye is looking) and other gaze-based phenomena \citep{duchowski2007eye}. As noted above, changes in pupil diameter can be due either to cognitive processing, in a phenomenon also known as task-evoked pupillary responses (TEPRs), or to other types of pupillary responses, such as that to ambient light \citep{Beatty2000}. The metrics most commonly employed to quantify TEPRs are \textit{mean pupil dilation}, \textit{maximum pupil dilation} (i.e., \textit{peak dilation}), and \textit{latency to peak dilation}. These metrics are computed about 0.5 to 5 seconds after initiation, with the time depending on the nature of the task \citep{Beatty1982}. For analysis of TEPRs under controlled lighting conditions it is common to divide the signal into epochs initially, in accordance with the nature of the task or the stimuli presented. For each of the epochs, baseline normalisation usually is required, in such a form as subtracting from the epoch pupil data for some period of time prior to presenting of the stimulus (e.g., 500~ms; see \citet{Beatty2000}), following which metrics can be computed.

\begin{figure}[!t]
   \centering
   \includegraphics[scale=0.5]{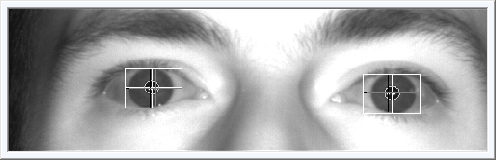}
   \caption{The pupil as measured through video-oculography.}
   \label{fig.Pupil}
\end{figure}

A large body of research has considered the relationship between the individual elements affecting pupil dilation. For instance, \citet{Palinko2011,Palinko2012} examined whether cognition-related pupil dilations can be separated from luminance-related pupil dilations in driving simulation studies. Although highly dependent on the experimental procedure, their results indicate that it is indeed possible to separate these two components of pupil dilation. More recently, \citet{Binda2014} have been studying how the interaction of luminance and cognitive load is reflected in pupil size. 

\section{Applications}
To illustrate the potential for pupil size metrics in HCI, we selected a subset of state-of-the-art, non-overlapping studies in which pupil size has been used on its own rather than in combination with other psychophysiological signals. 

As previously mentioned, one of the most well-established indices that can be inferred from pupil measurements is utilised for estimating mental workload. Examples of pupil size as an indicator of mental workload include the Index of Cognitive Activity (ICA), developed by \citet{Marshall2002}. In Marshall's work, pupil size is used for real-time estimation of the user's cognitive effort in the course of interaction with a virtual display. More recently, \citet{takeuchi2011} studied whether cognitive effort as reflected in pupil size changes as the perceptual learning process progresses in visual search tasks. They found that in the early stages of the learning process, pupil size rapidly increased in line with mental effort. However, in later phases of learning, the increase was much less pronounced. Recently, \citet{Pedrotti2014} showed the predictive power of pupil size in relation to stress situations. They built a classifier, using neural networks, and were able to predict the stress condition being experienced by a participant by assessing pupil size. Their system achieved prediction accuracies of up to 79\%.

Pupil size has been used also as an indicator of affective processing. \citet{Partala2003} measured pupil size while participants were exposed to auditory emotional stimuli. They found that pupil dilation was significantly greater when the subjects were exposed to emotional stimuli as compared with neutral stimuli. In addition, the effect was found to be more prevalent among female subjects. 

In HCI research, \citet{Oliveira2009} showed how pupil size could be useful in analysis of perceived relevance of Web search results. They studied the relevance of images and documents. Carrying out controlled laboratory experiments, they found pupil size to differ significantly when subjects were viewing relevant vs. non-relevant search results. The authors suggested that pupil size is best viewed as a delayed measure of interest, because relevant changes in the pupils were found around 400 to 500 milliseconds after the stimuli were shown. In other work, \citet{Jepma2011} investigated the relationship between pupil dilation and choice strategy (exploration of new choices vs. exploitation of a fixed choice) and were able to differentiate between the two choice strategies considered: pupil dilation was significantly greater in exploration scenarios than in exploitation scenarios.

\section{Conclusion}
Pupil size has been studied in cognitive sciences for more than a century and has been slowly introduced in the HCI field in the last few decades. Pupil size can be used to infer cognitive workload and mental effort in a reliable manner, and it also can be measured relatively inexpensively and unobtrusively. However, the extensive dependence on external factors such as ambient light conditions is slowing its progress beyond controlled laboratory experiments. In a promising development, recent research aimed at discriminating task-evoked pupillary responses from external-evoked pupillary responses has shown positive results, which augurs a productive future for pupillometry in HCI. This may well herald application of the signal finding its way into less controlled set-ups `in the wild'.

\bibliographystyle{plainnat}
\bibliography{ch7_pupil_bib}

\end{document}